\documentclass[a4paper]{article}
\usepackage{listings}
\usepackage{xcolor}
\usepackage{multirow}
\usepackage{jheppub}  
\usepackage{dcolumn}
\usepackage[english]{babel}
\usepackage[utf8x]{inputenc}
\usepackage[T1]{fontenc}
\usepackage{palatino}
\usepackage[normalem]{ulem} %riscar palavra --> \sout{texto riscado}

\usepackage{amsmath}
\usepackage{afterpage}
\usepackage{graphicx, subcaption}
\usepackage[colorinlistoftodos]{todonotes}
\usepackage[colorlinks=true]{hyperref}
\usepackage{makeidx}
\newcommand{\be}{\begin{equation}}  
\newcommand{\ee}{\end{equation}}  
\newcommand{\bea}{\begin{eqnarray}}  
\newcommand{\eea}{\end{eqnarray}}  
\makeindex
\begin{document}

\vspace*{1.2cm}

\thispagestyle{empty}
\begin{center}
%{\LARGE \bf Effects of the magnetic field on the production of pseudoscalar mesons in ultraperipheral heavy-ion collisions}
{\LARGE \bf Effects of the magnetic field on $\pi^0$  production  in ultraperipheral Pb-Pb collisions}

\par\vspace*{7mm}\par

{

\bigskip

\large \bf C.N. Azevedo$^{1}$, R. Fariello$^{1,2}$,  F.C. Sobrinho$^{1}$, F.S. Navarra$^{1}$}

\bigskip

{\large \bf  E-Mail: cazevedo@if.usp.br\\
                     fariello@if.usp.br\\
                     fcsobrinho@usp.br\\
                     navarra@if.usp.br}

\bigskip

{$^1$Departamento de F\'{\i}sica Nuclear, Instituto de F\'{\i}sica, Universidade de S\~{a}o Paulo,
Rua do Mat\~ao 1371 - CEP 05508-090,
Cidade Universit\'aria, S\~{a}o Paulo, SP, Brazil\\
$^2$ Departamento de Ci\^encias da Computa\c{c}\~ao, 
Universidade Estadual de Montes Claros, Avenida Rui Braga, sn, Vila Mauric\'eia, CEP 39401-089, 
Montes Claros, MG, Brazil.}

\bigskip

{\it Presented at the Workshop of Advances in QCD at the LHC and the EIC, CBPF, Rio de Janeiro, Brazil, November 9-15 2025}

%{ E-mail: 
%cbaldenegro@ku.edu}

\vspace*{15mm}

\end{center}
\vspace*{1mm}

\begin{abstract}

In this work, we study the effect of the magnetic field on the production of neutral pions in photon-photon interactions in ultraperipheral Pb-Pb collisions at the LHC. The calculation is performed within the equivalent photon approximation, including a magnetic-field dependence in the decay width $\Gamma(\pi^0\to\gamma\gamma)$, from which the corresponding production cross section is computed. We find that the reduction of the two-photon decay width in the presence of a strong magnetic field leads to a substantial reduction (by a factor of about 2-3) of the $\pi^0$ production cross section at LHC energies.

\end{abstract}

% \section{1st section}
 \section{Introduction}

Relativistic heavy-ion collisions are known to generate extremely intense magnetic fields, which have motivated extensive investigations of their impact on the properties of strongly interacting matter \cite{kharzeev2007, skokov2009, voronyuk2011, deng12, blocz13, guo20, kharzeev2013, adhikari2024}. Recent calculations indicate that the maximum intensity of the magnetic field generated in $Au+Au$ collisions at RHIC ($\sqrt{s}=200$ GeV) reaches values of the order of $eB \sim m_{\pi}^2 \sim 0.02$ $\text{GeV}^2$, while in $Pb+Pb$ collisions at the LHC ($\sqrt{s}=4.5$ TeV) even stronger fields can be produced, with intensities of up to $eB \sim 15 m_{\pi}^2 \sim 0.3$ $\text{GeV}^2$ \cite{kharzeev2007, skokov2009}. 
An important aspect of these extreme magnetic fields is their ability to induce anomalous transport phenomena in heavy-ion collisions, such as the chiral magnetic effect (CME) \cite{fukushima2008}, which predicts the preferential emission of charged particles along the direction of the angular momentum in systems with nonzero chirality. In addition, current theoretical investigations have focused on understanding how external magnetic fields modify hadronic properties, particularly in the nonperturbative regime of QCD. Since first-principles calculations in this regime are highly challenging, much of the current understanding relies on  effective approaches, such as chiral perturbation theory (ChPT) \cite{agasian2001}, quark-meson models \cite{ayala2020}, as well as Nambu-Jona-Lasinio (NJL) type models \cite{fayazbakhsh2012, avancini2015}. 
It is well known that strong magnetic fields, such as those generated in noncentral heavy-ion collisions, can significantly alter quark dynamics during the early stages of the interaction. As a consequence, the effects of these fields on light mesons, including their masses and decay constants, have attracted increasing attention \cite{andersen2012, bali2018, coppola2019, scocc25}.

Although neutral mesons do not couple directly to the external magnetic field, their internal quark structure makes them sensitive to the  field through quark-level interactions. In particular, the study presented in Ref.~\cite{scocc25} investigates the anomalous decay $\pi^0\to\gamma\gamma$ in the presence of intense magnetic fields within the two-flavor NJL model and demonstrates that the corresponding decay width is strongly reduced as the  field increases. 
Alternatively one can adopt a hadronic framework, as in \cite{shov25}, where the neutral pion decays through a proton triangular loop. 
The interaction between the neutral pion and the proton is described by the Yukawa-type Lagrangian: 
$ \mathcal{L} =i \, \lambda \,  \pi^0 \,  \bar{\psi}_p  \, \gamma^5 \,  \psi_p $ where $\lambda = m_p /f_{\pi}$. In the weak field limit both quark and 
hadron approaches should yield similar results. In this limit, over distance scales of the order of the magnetic length 
$ l = 1 / \sqrt{|e B|} $ the proton can be accurately treated as a point-like particle, and the details of its quark substructure are not important. On the other hand in the strong field limit the magnetic length may become very small and 
the quark structure may become relevant. In \cite{shov25}, as in \cite{scocc25}, the authors find that the qualitative effect of the magnetic field is to reduce the pion decay rate. However, unlike in \cite{scocc25}, they find that the effect is small for realistic field strengths. For magnetic fields 
of order $|e B| \approx m_{\pi}^2 \approx 0.02$ GeV$ ^2$,  typical of those expected in heavy-ion collisions, the correction to the decay rate is only about 
$0.02$ \% and even for $|e B| \approx 10 \, m_{\pi}^2 \approx 0.2$ GeV$ ^2$ the suppresion is only of 2 \%. 
The  origin of the differences between the quark and the hadron approach is not yet  understood, but they probably come from different model assumptions and,
in particular, from the treatment of the proton and the pion as  point-like particles. In this work we will explore the consequences of  a strong reduction  
in the pion decay width found in  \cite{scocc25}. 

Photon-induced processes have been intensively studied in ultraperipheral collisions (UPCs) for several years \cite{bertulani1988,bertulani2005}, providing 
an excellent environment to investigate effects discussed here. In UPCs, the nuclei interact with a large impact parameter, such that no nuclear overlap occurs. 
Typically $b > 2R$, where  $R$ is the radius of the nucleus, allowing the interaction to be dominated by the intense electromagnetic fields surrounding the ions. Moreover, nuclei with large nuclear charge $Z$ can provide an ideal testing ground for the study of photon-induced processes, due to the huge number of coherent photons associated with the highly Lorentz contracted electromagnetic fields ($\propto Z^2$).  As in non-central heavy ion collisions, a strong magnetic field is also created in UPCs. In this clean environment, the effects of the $B$ field are expected to be much easier to observe than in non-central collisions. In \cite{isa1,isa2} it was shown that in UPCs the magnetic field induces $N \to \Delta$ transitions in the projectiles with very large cross sections.  The delta resonance then decays into pions, which have momenta very close to the beam momentum.  If the forward pion is a $\pi^0$ it can be detected through its decay into two photons. In principle, the measurement of these forward photons can be performed with the detectors used by the LHCf \cite{lhcf} 
collaboration and would allow us to indirectly  measure the magnetic field.

In the present work, we investigate the effects of an intense magnetic field on neutral pion production in photon-photon interactions in ultraperipheral heavy-ion collisions. In particular, we analyze the impact of the magnetic-field-dependent $\pi^0\to\gamma\gamma$ decay width on the corresponding production cross section. Our results indicate that an intense background magnetic field can, in principle, lead to a  reduction by a factor of about 2-3 in the pion production cross section  at LHC energies.

%\section{2nd section}
\section{Formalism}

\begin{figure}[t!]
    \centering
    \includegraphics[width=.7\linewidth]{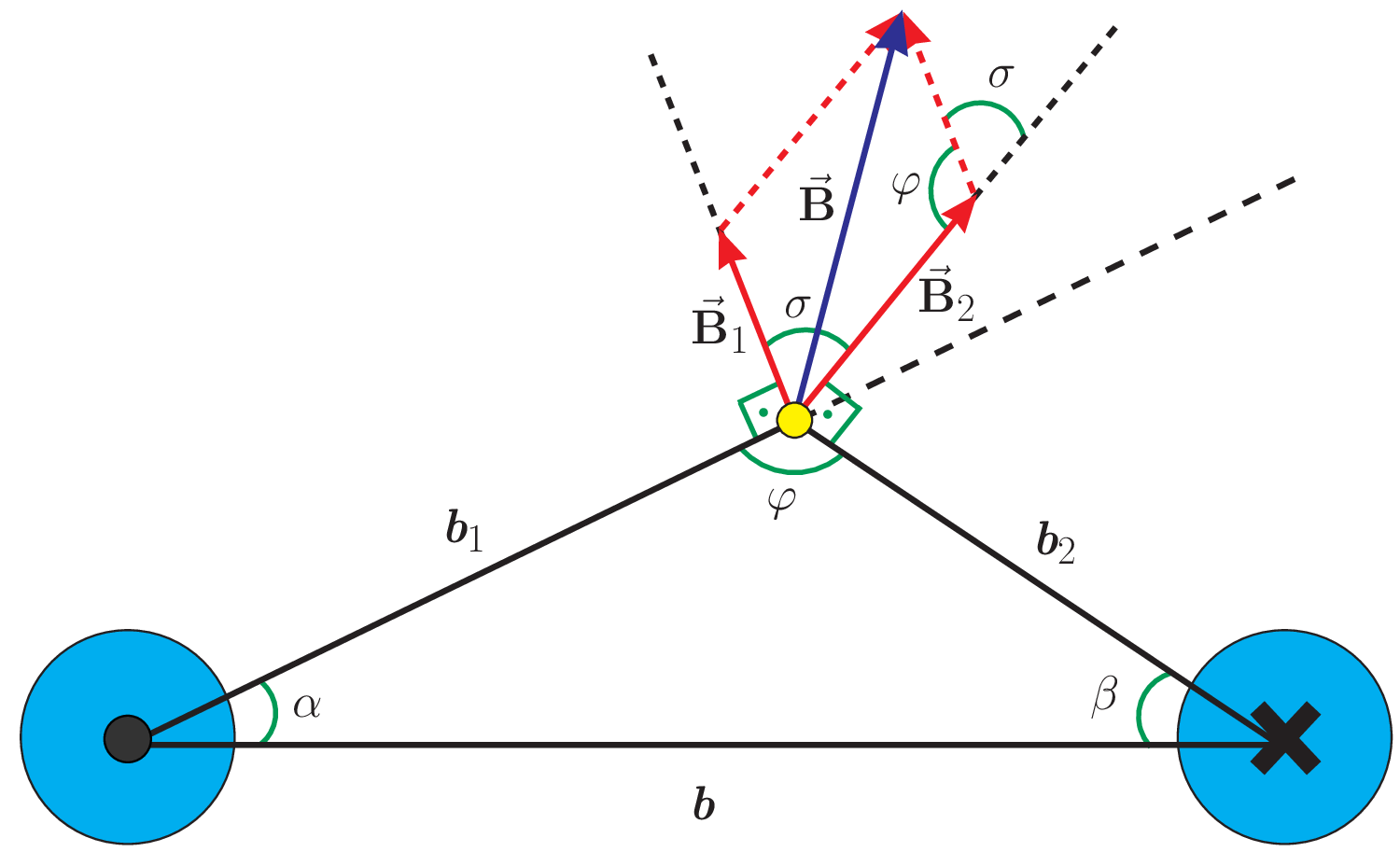}
    \caption{The circles represent one nucleus coming out of the page (left) and the other going into the page (right) generating the magnetic fields
    $\vec{B}_1$ and $\vec{B}_2$ respectively. Two photons interact at the yellow circle, where the $\pi^0$ is formed and feels the field $\vec{B}$.  
    } 
    \label{fig:parameter-b} 
\end{figure}

In this section, we present a brief review of the main concepts needed to describe the $\pi^0$ production in $\gamma\gamma$ interactions in $PbPb$ collisions. In the equivalent photon approximation (EPA), the total cross section is given by \cite{budnev1975} 
\begin{equation}
    \sigma(PbPb \to Pb \otimes \pi^0 \otimes Pb; s) = 
    \int 
    \hat{\sigma}(\gamma\gamma \to \pi^0; W)
    N(\omega_1,\mathbf{b}_1) N(\omega_2,\mathbf{b}_2) 
    S_{abs}^2(\mathbf{b}) 
    \mathrm{d}^2 \mathbf{b}_1 \mathrm{d}^2 \mathbf{b}_2
    \mathrm{d}\omega_1 \mathrm{d}\omega_2  
    ,
\label{eq:cross-section}
\end{equation} 
where $\sqrt{s}$ is the center-of-mass energy of the $PbPb$ collision, the symbol $\otimes$ represents the presence of a rapidity gap in the final state, and $W=\sqrt{4\omega_1\omega_2}$ is the invariant mass of the $\gamma\gamma$ system. The term $N(\omega_i, \mathbf{b}_i)$ represents the equivalent photon spectrum generated by nucleus $i$, which determines the number of photons with energy $\omega_i$ at a transverse distance $\mathbf{b}_i$ from the center of the nucleus, defined in the plane perpendicular to the trajectory, as shown in Fig. \ref{fig:parameter-b}. The quantity $\hat{\sigma}(\gamma\gamma \to \pi^0)$ is the photoproduction cross section of $\pi^0$ from two real photons with energies $\omega_1$ and $\omega_2$. Moreover, the factor $S^2_{abs}(\mathbf{b})$ is the absorption factor, responsible for excluding the overlap between the nuclei and ensuring that only ultraperipheral collisions are taken into account, and is given by \cite{baur1990} 
\begin{equation}
    S_{abs}^2(\mathbf{b}) = \Theta(|\mathbf{b}|-R_1-R_2) 
    \,=\, 
    \Theta(|\mathbf{b}_1 -\mathbf{b}_2|-R_1-R_2) 
    \,,
\label{eq:abs}
\end{equation}
where $R_i$ is the radius of nucleus i. The relevant quantities for the calculation of the impact parameter are illustrated in Fig. \ref{fig:parameter-b}, which shows a view in the plane perpendicular to the direction of motion of the two nuclei. The parameter $\mathbf{b}$, which characterizes the collision, is related to the distances $\mathbf{b}_1$ and $\mathbf{b}_2$ between the centers of nucleus 1 and 2  and the interaction point through the relation $b^2 = b_1^2+b_2^2 - 2b_1b_2\cos \, \varphi$. The equivalent photon flux can be expressed as \cite{krauss1997}, 
\begin{equation}
N(\omega_i, b_i) =  
\frac{Z^2 \alpha}{\pi^2}\,\frac{1}{b_i^2 v^2 \omega_i} 
\left[ \int_{0}^{\infty} u^2 \,  J_1(u) \, F \left( \sqrt{\frac{u^2 + (b_i \, \omega_i / \gamma_L)^2}{b_i^2} } \, \right)
                        \frac{1}{u^2 + (b_i \, \omega_i / \gamma_L)^2 } \; du
\right]^2 
\,,
\label{eq:flux}
\end{equation}
where $\alpha$ is the electromagnetic coupling constant, $J_1$ is the Bessel function of the first kind, $\gamma_L$ is the Lorentz factor, and $v$ is the velocity of the nucleus, assumed to be equal to $c$. Moreover, $F$ is the nuclear form factor of the equivalent photon source. In our analysis, we will employ a realistic form factor associated with the Woods–Saxon distribution, obtained from the Fourier transform of the nuclear charge density and constrained by experimental data. This form factor can be analytically approximated by \cite{bertulani2001}
\begin{equation}
F(q^2) = \frac{4\pi\rho_0}{A \, q^3} \, 
[\sin(q\, R) - q \, R \, \cos(q\, R)] 
\left[
    \frac{1}{1+q^2 a^2}
\right] 
\,,
\label{eq:realistic}
\end{equation}
with $R=6.63$ fm, $a = 0.549$ fm, and $\rho_0 = 0.1604$ $\text{fm}^{-3}$ for the lead nucleus \cite{DeJager1974, DeVries1987}. An alternative way to describe the nuclear charge distribution is to employ a monopole form factor, given by: 
\begin{equation}
    F(q^2) = 
    \frac{\Lambda^2}{\Lambda^2 + q^2}
    \,, 
\label{eq:monopole}
\end{equation}
where $\Lambda$ is a parameter adjusted to reproduce the root-mean-square (rms) radius of a nucleus. For the case of $^{208}Pb$, we have $\Lambda = 0.088$ GeV \cite{DeVries1987}. Considering the Low formula \cite{low1960}, the cross section $\hat{\sigma}(\gamma\gamma\to\pi^0)$ can be written in terms of the two-photon decay width of the $\pi^0$ ($\Gamma_{\pi^0\to\gamma\gamma}$) as 
\begin{equation}
    \hat{\sigma}_{\gamma\gamma\rightarrow \pi^0}(\omega_1,\omega_2) = 
    8\pi^2 (2J + 1) \,
    \frac{\Gamma_{\pi^0  \to \gamma\gamma}}{M_{\pi^0}} \,
    \delta(4\omega_1\omega_2 - M_{\pi^0}^2) 
    \,,
\label{eq:low}
\end{equation}
where $J$ and $M_{\pi^0}$ are, respectively, the spin and the mass of the produced neutral pion. In the present calculation, we assume that the pion mass does not depend explicitly on the magnetic field, while the partial decay width of the $\pi^0$,  $\Gamma_{\pi^0 \to \gamma\gamma}$,  does.    
In our analysis, we will consider the effect of the magnetic field on the decay width and, consequently, on the total cross section. 
Detailed calculations of these magnetic fields depend on a series of assumptions. For instance, it is common to neglect the contribution  from particles produced in the collision. Therefore, it is sufficient to take into account only the colliding particles. In the center-of-mass frame, the two nuclei represent electric currents flowing in opposite directions which, according to Maxwell’s equations, generate the magnetic fields $\vec{B}_1$ and $\vec{B}_2$. As depicted in Fig. \ref{fig:parameter-b},  at a given point the resulting
field is given by the vector sum of $\vec{B}_1$ and $\vec{B}_2$ and its absolute value is given by:
\begin{equation}
    B^2 = B_1^2 + B_2^2 - 2 \, B_1 \, B_2 \, cos  \, \varphi
    \,, 
\label{bfinal}
\end{equation}
The magnetic field depends on the ion energy, the impact parameter $b$ of the collision, as well as on the position and time. 
For simplicity, we assume that the magnetic field is the same as that generated by a point charge in motion. Its absolute value is expressed as 
\cite{asakawa2010, isa1,isa2}:
\begin{equation}
    B_i = \frac{1}{4\pi}\,
    \frac{Ze\,\gamma_L b_i\, v}{[\gamma_L^2 v^2 t^2 +  b_i^2]^{3/2}} 
    \,, 
\label{magnetic-field}
\end{equation}
which corresponds to the general time-dependent magnetic field at a transverse distance $b_i$ from the trajectory of the nucleus i ($i=1,2$). 
As it can be seen, this field depends on time. To consistently take into account the time dependence of our formalism we would need to rewrite our main
formula, Eq. (\ref{eq:cross-section}), with a time dependent photon flux  $N(\omega_i, b_i,t)$. This would require a deeper reformulation of the EPA 
formalism, which is beyond the scope of this work. 
The magnetic fields in Eq. (\ref{magnetic-field}) have a maximum at $t=0$, i.e. when the two nuclei (treated as two extremely Lorentz contracted 
"sheets") are in the plane transverse to the reaction plane. We will assume that the reaction occurs only at this instant and hence under the 
influence of the magnetic field at $t=0$. 

\noindent 
Taking $t=0$ in Eq. (\ref{magnetic-field})  and multiplying both sides by the elementary charge $e$ (where $e=\sqrt{4\pi\alpha}$) we obtain 
\begin{equation}
    eB_i = \frac{Z\alpha\,\gamma_L v}{b_i^2}
    \,.
\label{eB}
\end{equation}
When the magnetic field reaches intensities comparable to typical QCD energy scales, such as $eB \sim m_{\pi}^2$, its effects become dynamically relevant, requiring that the quark structure be taken into account. The calculation of the neutral pion decay width into two photons, taking into account its dependence on an external magnetic field, can be performed within effective theories or through LQCD calculations. In Ref. \cite{scocc25}, the dependence on the magnetic field was incorporated using the SU(2) version of the local NJL model. In what follows, we will parametrize the results obtained in \cite{scocc25} and 
find  a  simple expression for the $B$-dependent decay width, $\Gamma_{\pi^0 \to \gamma\gamma}(B)$.

\section{Results and discussion}

\begin{figure}[t]
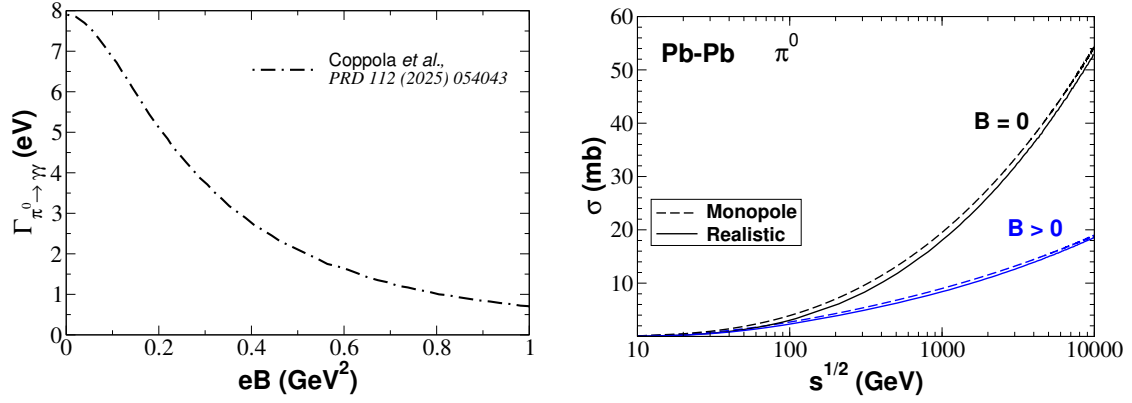

\begin{tabular}{cc}
\includegraphics[width=.458\linewidth]{./coppola.eps} & \;
\includegraphics[width=.475\linewidth]{./energy_logx.eps}  
%\\
%  (a) & (b)
\end{tabular}
\caption{Left: $\Gamma_{\pi^0 \to \gamma\gamma}$ as a function of $eB$ for lowest-order chiral expansion, from Ref. \cite{scocc25}. 
Right: energy dependence of the total cross section for $\pi^0$ production in $\gamma\gamma$ interactions in ultraperipheral $PbPb$ collisions. The black curves show the results for a zero magnetic field, while the blue curves correspond to a nonzero magnetic field. The solid (dashed) line correspond to calculations using the realistic (monopole) form factor.}
\label{fig:energy}
\end{figure}
In Fig. \ref{fig:energy} (left),  we present numerical results from Ref. \cite{scocc25} for the decay width of the neutral pion into two photons as a 
function of the magnetic field, accounting for magnetic-field effects on sea quarks through an effective field-dependent four-quark interaction. 
As it can be seen,  the total decay width is strongly reduced in the presence of the external field. 
In Fig. \ref{fig:energy} (right), we present our predictions for the energy dependence of the total cross section for $\pi^0$ production in ultraperipheral $PbPb$ collisions. The black curves correspond to the case in which the magnetic field is set to zero, while the blue curves incorporate the variations in 
the value of the decay width $\Gamma(\pi^0\to\gamma\gamma)$ induced by the magnetic field computed with Eqs. (\ref{bfinal}),  (\ref{magnetic-field}) and
(\ref{eB}).  The solid (dashed) lines were obtained using the realistic (monopole) form factor.

\noindent 
The predictions for the cross sections of neutral pion production in $PbPb$ collisions at $\sqrt{s}=5.02$ TeV were obtained using the formalism described previously, with the realistic form factor for the photon flux. For the case without a magnetic field ($B=0$), we obtained 40 mb. This value is in agreement with previous estimates  based on the equivalent photon approximation \cite{navarra2023}. On the other hand, when the effects of the magnetic field ($B>0$) are included, the total cross section is estimated to be 15 mb, indicating a significant reduction associated with the strong suppression of the decay width $\Gamma(\pi^0\to\gamma\gamma)$ induced by the magnetic field. This reduction by a factor of about 2 to 3 indicates that the effects of the magnetic field may have a measurable impact on the production of light mesons in UPCs.

%\section{Conclusions}

%The strong suppression of the decay width $\Gamma(\pi^0\to\gamma\gamma)$ induced by the magnetic field leads to a considerable reduction in the predicted cross section for $\pi^0$ production. This result indicates that intense magnetic fields can modify photon–photon processes in heavy-ion collisions at a level that may, in principle, be experimentally relevant. Therefore, the investigation of magnetic field effects on $\pi^0$ production in ultraperipheral collisions is strongly motivated, especially considering the possibility of measuring such processes in experiments at the LHC. 
%
\noindent 
 The present analysis was performed under simplified assumptions for the magnetic field, in particular considering the magnetic field evaluated only at its maximum value ($t = 0$). A more realistic description should take into account the time dependence of the equivalent photon flux and 
 of the magnetic field generated in relativistic heavy-ion collisions. The study of these time-dependent effects will be the subject of future work. Nevertheless, the magnitude of the effects observed in the present study suggests that the contributions of the magnetic field may be sufficiently large to justify their inclusion in more detailed simulations of heavy-ion collisions.

\section*{Acknowledgements}

This work was partially financed by the Brazilian funding agencies CNPq, CAPES, FAPESP and INCT-FNA (Process No. 464898/2014-5). C.N. Azevedo gratefully acknowledges the support from the Funda\c{c}\~ao de Amparo \`a  Pesquisa do Estado de S\~ao Paulo (FAPESP), Process No. 2023/14456-8. R. Fariello received funding from CNPq (grant 173012/2023-0). F.C. Sobrinho gratefully acknowledges the support from the Funda\c{c}\~ao de Amparo \`a  Pesquisa do Estado de S\~ao Paulo (FAPESP), Process No. 2024/19103-9.

\end{document}